\documentclass[review,11pt]{article}
\usepackage{times}
\usepackage{soul}
\usepackage[normalem]{ulem}
\usepackage[squaren,Gray]{SIunits}
\usepackage{empheq}
\usepackage{bm}
\usepackage{epstopdf}
\usepackage{subfigure}
\usepackage{color}
\usepackage[colorlinks=true,citecolor=blue,linkcolor=red]{hyperref}
\usepackage{booktabs}

\usepackage[natbibapa]{apacite}

\renewcommand{\d}{{\rm d}}			

\usepackage[tablesfirst,nomarkers]{endfloat}
\begin{document}
\title{Numerical modelling of open channel junctions using the Riemann problem approach}
\author{MOHAMED ELSHOBAKI, PhD,\\
\textit{Department of Information Engineering, Computer Science, and Mathematics,}\\
\textit{University of L'Aquila, L'Aquila, Italy} \\ 
\textit{Email: \textcolor{blue}{mohabd@univaq.it}} 
\and
ALESSANDRO VALIANI (IAHR Member), Professor,\\
\textit{Department of Engineering, University of Ferrara, Ferrara, Italy} \\
\textit{Email: \textcolor{blue}{alessandro.valiani@unife.it} (author for correspondence)}
\and
VALERIO CALEFFI, Assistant Professor,\\
\textit{Department of Engineering, University of Ferrara, Ferrara, Italy} \\
\textit{Email: \textcolor{blue}{valerio.caleffi@unife.it}}}
\maketitle

\newpage
\noindent \textbf{Numerical modelling of open channel junctions using the Riemann problem approach}\\ \\ \\
\begin{abstract}
The solution of an extended Riemann problem is used to provide the internal boundary conditions at a junction when simulating one-dimensional flow through an open channel network. The proposed approach, compared to classic junction models, does not require the tuning of semi-empirical coefficients and it is theoretically well-founded. The Riemann problem approach is validated using experimental data, two-dimensional model results and analytical solutions. In particular, a set of experimental data is used to test each model under subcritical steady flow conditions, and different channel junctions are considered, with both continuous and discontinuous bottom elevation. Moreover, the numerical results are compared with analytical solutions in a star network to test unsteady conditions. Satisfactory results are obtained for all the simulations, and particularly for Y-shaped networks and for cases involving variations in channels' bottom and width. By contrast, classic models suffer when geometrical channel effects are involved. 
\end{abstract}
\section{Introduction}
\label{introduction}
Channel junctions are found in natural rivers, irrigation and drainage canals, and urban wastewater networks. Therefore, understanding such systems is an essential issue in Hydraulics, where the computation of the water surface profiles is necessary for both steady and unsteady flows. When the water depth is sufficiently small compared to the typical horizontal scale, as in river and channel networks, one-dimensional (1D) St. Venant equations (in which cross-sectional area and total discharge are the main variables) are widely used to describe the flow \citep{chow1959}; under the assumption of rectangular cross-section, the St. Venant model matches the one-dimensional shallow water equations (SWEs).

The 1D SWEs are solved by means of different numerical methods, such as the finite difference method (FDM), finite element method (FEM), and finite volume method (FVM); see \citep{briani2016notes,unami2012concurrent,
bellamoli2017numerical,neupane2015discontinuous,
aral1998application,kesserwani2008simulation,
ghostine2012comparative,
ghostine2010two,borsche2016numerical} and the references therein.

Independently from the specific adopted numerical scheme, using the 1D approach to numerically solve open channel networks faces mathematical difficulties at the intersection of the channels (i.e., junctions). 
Indeed, whilst in a 2D framework the numerical simulation of a junction doesn't require particular precautions, in a 1D framework the junction is a singular point, where the numerical scheme cannot be directly applied and therefore internal boundary conditions must be prescribed. The system of governing equations used to supply the internal boundary conditions must have a solution and this solution must be unique \citep{JHDE2018}. Moreover, a proper numerical treatment of these boundary conditions is required to ensure the well-posedness of the numerical scheme \citep{colombo2006well}. 

Considering only subcritical flows, which are the most common in nature, the well-established methods to construct the internal boundary conditions are based on four classic approaches. The first approach is reported by \citet{akan1981diffusion}, which prescribes that the total energy is preserved at junctions, being approximated by the water depth, while kinetic head is neglected. The second approach is introduced by \citet{Karki1997} and considers the momentum balance together with the mass conservation applied at the junction. The third approach \citep{JungLee1998} extends the principles given in \citet{Karki1997} introducing energy and momentum coefficients to include the energy losses at the junction. The reader is addressed to \citep{pinto2015experimental,leite2012flow} for a thorough discussion on the physics behind this approach. Finally, \citet{Shabayek2002} use a general nonlinear formulation of the momentum principle and the conservation of mass, which remove the restriction of equality of channel depths and channel widths at the junction.
The equations associated to these approaches are coupled to the continuity equation \citep{chow1959} and the characteristic equations \citep{ABBOTT1966,chaudhryopen} to form the six-equation, nonlinear system governing the junction \citep{JHDE2018}.
We refer to the formulations associated to these four approaches as the Equality model \citep{akan1981diffusion}, Gurram model \citep{Karki1997}, Hsu model \citep{JungLee1998}, and Shabayek model \citep{Shabayek2002}, respectively. 
Note that the Shabayek model implies the use of two empirical coefficients that require further characterization, as stated in \citet{pinto2015experimental}, and it is therefore excluded from the present analysis. 

A study by \citet{kesserwani2008simulation}, comparing Hsu, Gurram, Equality and Shabayek models for subcritical junction flow, shows that the Equality model leads to poor momentum conservation when the Froude number is greater than $0.35$. The study also finds that the influence on the flow of the angle between the main and lateral channels is much less important than the Froude number downstream of the junction. However, the results are only presented for a specific type of asymmetric confluence \citep{best1985flow}, under the assumption of a flat bottom throughout the junction. Other works on the topic are carried out considering further comparison with 2D results \citep{ghostine2009confrontation}, or supercritical and transcritical flows \citep{kesserwani2008one, kesserwani2010new}.

The classic methods are not tested in symmetric confluences (i.e., Y-shaped) because such methods, and particularly the Gurram and Hsu models, are not derived for this type of confluence. The flow field at confluences is also affected by bottom discordance (i.e., a bottom discontinuity at the confluence) between the lateral and mainstream channels \citep{best1988sediment,bradbrook2001role,biron1996effects,wang2007three,leite2012flow}. 
To extend the Gurram and Hsu models to Y-shaped confluences, the fundamental governing equations are re-derived in this work, taking into account the differences in geometry and bottom elevation.

As an alternative to the classic methods, a recent formulation of the internal boundary conditions is proposed by \citet{briani2016notes}, based upon the work by \citet{goudiaby2013riemann}. This formulation is obtained by solving a well-posed Riemann problem (RP) at the junction assuming a continuous bottom and symmetric configurations. We refers to this formulation as the Riemann problem approach (RP approach). A rigorous study about the existence and uniqueness of the problem solution is also provided for the symmetric case without bottom steps \citep{goudiaby2013riemann}. The RP approach is theoretically analyzed in more general configurations by \citet{JHDE2018}, where asymmetric networks and discontinuous bottom are taken into account.

The purpose of this work is to compare the RP approach with the classic approaches. In particular, we are interested in the application aspects. 
With this aim, the classic junction models and the RP approach are implemented in a FVM Dumbser-Osher-Toro (DOT) scheme \citep{dumbser2011simple}.
The models are tested against experimental data provided in literature \citep{JungLee1998,hsu1998flow,briani2016notes,bradbrook2001role,wang2007three,biron1996effects} for steady  flows in both asymmetric and symmetric confluences. In particular, only discordant bottoms are considered in the experimental data of \citet{bradbrook2001role,wang2007three,biron1996effects}.
To complete the current study, the models are tested against the analytical solutions provided by \citet{goudiaby2013riemann} for unsteady flows.

The rest of this paper is structured as follows. First, the mathematical model and its numerical treatment are given. Then, the Riemann, Equality, Gurram, and Hsu junction models are briefly described. Next, the models are tested for both steady and unsteady open channel flows, and numerical results are presented. The numerical solutions are compared with the experimental results and analytical solutions. Finally, conclusions are given.  

\section{Mathematical model and numerical scheme}
\label{1dswe}
In this section, the shallow water equations are described. Then, the FVM-DOT numerical scheme \citep{dumbser2011simple}, used to discretize the SWEs in each channel, is briefly outlined.
\subsection{The one-dimensional shallow water equations}
The SWEs are a particular case of the Navier-Stokes equations and are obtained by integrating the mass and momentum equations for an incompressible fluid over the depth. They are written in conservative form as:
\begin{equation}
\label{eq1}
\frac{\partial{\bm{U}}}{\partial {t} }+ \frac{\partial {\bm{F}}}{\partial {x}} = \bm{S},\,\,\,\,  \textrm{in}\, [0,L]
\end{equation}
with 
\begin{align*}
\bm{U}&=\begin{bmatrix} h\\ hu \end{bmatrix},& \bm{F}&=\begin{bmatrix} hu\\ hu^2 +\frac{{g}h^2}{2} \end{bmatrix},& \bm{S}&=\begin{bmatrix}0 \\ {g}h(S_{0x}-S_{f})\end{bmatrix},
\end{align*}
where $u(x,t)$ and $h(x,t)$ are the flow velocity and the flow depth, respectively. $L$ is the channel length; ${g}$ is the gravity acceleration; $S_{0x}=-\partial z/ \partial x$ is the bottom slope; $z(x)$ is the bottom elevation; $S_{f}$ is the friction slope \citep{chow1959}; and $x$ and $t$ are space and time, respectively. For the purpose of this study, the forces due to friction are much smaller than pressure forces and momentum fluxes, so the SWEs are solved as in the frictionless bottom case (i.e., $S_{f}=0$). Eq. (\ref{eq1}) can therefore be cast in a quasi-linear form as follows:
\begin{equation}
\label{eq2}
\frac{\partial{\bm{W}}}{\partial {t} }+ \mathbf{A}(\bm{W})\frac{\partial {\bm{W}}}{\partial {x}} = 0, \, \, \, \textrm{in} \, [0,L]
\end{equation}
with
 \[ \bm{W}=\begin{bmatrix} h\\ hu\\z \end{bmatrix}, \quad \mathbf{A}(\bm{W})=\begin{bmatrix} 0 & 1 &0 \\ gh-u^{2} & 2u  &gh \\ 0& 0& 0  \end{bmatrix}.\]

The form of the SWEs in Eq. (\ref{eq2}) is preferable when bottom discontinuities have to be included in the mathematical model \citep{lefloch2007riemann,caleffi2017well,valiani2017momentum}. This aspect is fundamental because a discontinuity in bottom elevation is a recurring feature at junctions \citep{leite2012flow}.
\subsection{{Dumbser-Osher-Toro Riemann solver}}
\label{numerical}
The integration of Eq. (\ref{eq2}) over a control volume gives the following path-conservative formulation  \citep{pares, dumbser2011simple}:
\begin{equation}
\label{eq5}
\bm{W}_{i} ^{n+1}= \bm{W}_{i} ^{n}- \frac{\Delta {t}}{\Delta {x}}(\bm{D}_{i+\frac{1}{2}}^{-} +\bm{D}_{i-\frac{1}{2}}^{+}),
\end{equation}
where the fluctuations $\bm{D}_{i\pm\frac{1}{2}}^{\pm}$ must satisfy the following compatibility condition:
\begin{equation}
\label{eq6}
\bm{D}_{i+\frac{1}{2}}^{-} +\bm{D}_{i+\frac{1}{2}}^{+}= \int_{0}^{1} \mathbf{A}(\mathbf{\psi} (\bm{W}_{i+1},\bm{W}_{i},s))\frac{\partial{\mathbf{\psi}}}{\partial {s}} \d s.
\end{equation}
$\bm{W}_{i}^{n}$ denotes the cell average of the conservative variables at time $t^n$. The uniform spatial step is $\Delta{x}=x_{i+\frac{1}{2}}-x_{i-\frac{1}{2}}$ and the time step $\Delta{ t}= t^{n+1}-t^n$. Choosing a linear integration path $\mathbf{\psi}(s)$  \citep{dumbser2011simple} in the parameter $s \in [0,1]$:
\begin{equation}
\label{eq8}
\mathbf{\psi}(s)= \mathbf{\psi}(\bm{W}^-, \bm{W}^+,s)=\bm{W}^- + s(\bm{W}^+ - \bm{W}^-)
\end{equation}  
the Osher fluctuation term becomes:
\begin{equation}
\label{eq9}
\bm{D}_{i+\frac{1}{2}}^\pm = \frac{1}{2}\bigg(\int_{0}^{1} \mathbf{A}(\mathbf{\psi}(s)) \pm |\mathbf{A}(\mathbf{\psi}(s))|\,\d s \bigg)(\bm{W}_{i+1}  - \bm{W}_{i}).
\end{equation}
Eq. \eqref{eq9} is replaced by
\begin{equation}
\label{eq10}
\bm{D}_{i+\frac{1}{2}}^\pm = \frac{1}{2}\bigg( \sum_{j=1}^{G}\omega_{j}\left[\mathbf{\mathbf{A}(\psi}(s_{j}))\pm |\mathbf{A}(\mathbf{\psi}(s_{j}))|\right]\bigg)(\bm{W}_{i+1} - \bm{W}_{i}),
\end{equation}
using a $G$-point Gauss-Legendre quadrature in the interval $[0,1]$ with nodes $s_j$ and weights $\omega_j$ \citep{stroud1971}. 
For the stability of the scheme, the time step must satisfy the relationship: 
\begin{equation}
\label{eq11}
\Delta t={\textrm{CFL}} \frac{\Delta x}{{\rm{max}}(| u \pm c |)},
\end{equation}
where $\textrm{CFL}<1$ is the Courant-Fredrich-Lewy coefficient, and $c=\sqrt{{g}h}$ is the wave celerity.

Finally, the scheme has to be completed with boundary conditions. Two types of boundary conditions are needed: external and internal. The external boundary conditions are posed at the inflow-outflow sections of the network. They are defined by taking into account the subcritical flow state considered in this work. A discharge hydrograph is imposed at the inflow sections and a given water depth is imposed at the outflow sections. The external boundary conditions are numerically treated as described by \citet{chaudhryopen}.

The internal boundary conditions are imposed at the interfaces between the channels at the junction node. At the extremity of each channel adjoining the node, a depth and a discharge must be prescribed. Therefore, for a network of three channels, the unknowns are three water depths and three water discharges. To compute these unknowns, a junction model which takes shape of a system of six equations must be given. In Section \ref{junction}, the junction models considered in this work are briefly summarized.
\section{Junction models}
\label{junction}
This section presents a short description of the nonlinear junction models used here to provide the internal boundary conditions. Note that the classic Gurram and Hsu models are modified to include the effect of the lateral bottom discordance. In addition, these models are generalized to the case of a non-straight main channel in the Y-shaped confluence (see appendices \ref{appand1} and \ref{appand2}). 
\subsection{Riemann problem approach model}
The Riemann problem at the junction is defined by analogy as the classic Riemann problem in a single open channel. The classic Riemann solution has been described in \citet{Toro2009} for continuous bottom and in \citep{bernetti2008exact,alcrudo2001exact,lefloch2007riemann} for discontinuous bottom. Here, the Riemann problem consists of Eq. (\ref{eq2}) and the following initial conditions (depth, velocity and bottom elevation are assumed uniform in each channel):
\begin{eqnarray}
\label{initial_conditions}
\begin{cases}
h(x,0) = h_{0k} \\
u(x,0) = u_{0k} \\
z(x,0) = z_{k}
\end{cases}\quad k=1,2,3
\end{eqnarray}
where $k=1$, $k=2$, and $k=3$ refer to the main upstream channel, the lateral channel, and the main downstream channel, respectively. The unknowns  at the junction node can be predicted using the Riemann solution, as reported in \citep{goudiaby2013riemann}. The structure of the solution of the Riemann problem gives the following system:  
\begin{subequations}
\label{eq14}
\begin{equation}
\sum_{k=1} ^{3} \eta_{k}b_{k} h _{k} u_{k} =0 
\label{eq14:1}%
\end{equation}
\begin{equation}
\frac{u_{1}^2}{2{{g}}}+h _{1}+z_{1}=\frac{u_{k}^2}{2{{g}}}+h _{k}+z_{k},\quad k=2,3
\label{eq14:2}%
\end{equation}
\begin{equation}
u_{k}- u_{0k}+ \eta_{k} f(h_{0k},h_{k})=0,\quad k=1,2,3, 
\label{eq14:3}%
\end{equation}
\end{subequations}
where
\begin{equation}
\label{eq15}
f(h_{0k},h_{k})=
\begin{cases}
2(\sqrt{{{g}}h_{0k}}-\sqrt{{{g}}h_{k}}), \quad h_{k} < h_{0k}\\
(h_{0k}-h_{k})\sqrt{\frac{{g}}{2}( \frac{1}{h_{0k}} +\frac{1}{h_{k}})},\quad h_{k} \geq h_{0k}.
\end{cases}
\end{equation}
$(h_{0k}, u_{0k}) $ are the initial data and $b$ is the channel width. In the current work, $z_{1}=z_{3}=0$ and $z_{2}\neq 0$ to represent a bottom step between the second (lateral) channel and the main channels, as shown in Fig. \ref{f1}. 
The RP approach can be generalized for different bottom and junction configurations; the interest is here focused just on this case because it is the most frequent in natural streams \citep{leite2012flow}. The quantity
\begin{eqnarray}
\eta_{k}= 
\begin{cases}
\nonumber
\phantom{-}1,\quad \text{if} \quad x_{k}= {L}_{k},\quad k=1,2,3\\
-1,\quad \text{if}\quad x_{k}=0,\quad k=1,2,3
\end{cases}
\end{eqnarray}
refers to the inner boundary edge at the junction. Eq. (\ref{eq14:3}) represents the classic SWE wave relationships for shocks and rarefactions in each channel. In facts, Eq. (\ref{eq14:3}) is the Rankine-Hugoniot condition or the constancy of the Riemann invariants \citep{Toro2009} written in a convenient form. The continuity equation (\ref{eq14:1}) must be satisfied together with the equality of the total head at the junction Eq. (\ref{eq14:2}). The hypothesis of total head and flow discharge preservation in the 1D single channel over a bottom step, as part of the solution of the Riemann problem, is discussed in \citet{valiani2017momentum} and for the junction network of three non-identical channels in \citet{JHDE2018}.
\subsection{Equality model}
The equality model is the simplest junction model, and it is written in the following form:
\begin{subequations}
\label{eq17}
\begin{equation}
\sum_{k=1}^{3} \eta _{k} b_{k} h _{k} u _{k}=0 
\label{eq17:1}%
\end{equation}
\begin{equation}
h_{1}=h_{2}+z_{2}
\label{eq17:2}%
\end{equation}
\begin{equation}
h_{2}+z_{2}=h_{3}
\label{eq17:3}
\end{equation}
\begin{equation}
u_{k} h_{k}=A_{k}h_{k} +C_{k},\quad k=1,2,3,
\label{eq17:4}%
\end{equation}
\end{subequations}
where $A_{k}=u_{0k}\pm \sqrt{{{g}}h_{0k}}$ and $C_{k}=\mp h_{0k}\sqrt{{{g}} h_{0k}}$. The sign depends on the characteristic direction at the junction. Indeed, Eq. (\ref{eq17:1}) represents mass conservation. Eqs. (\ref{eq17:2}) and (\ref{eq17:3}) represent the equal water elevation condition at the junction, which was recognized by \citet{akan1981diffusion} as a simplification of the equal energy condition at the junction, where the kinetic head is considered to be small for subcritical flows. Eq. (\ref{eq17:4}) represents the characteristic equations, in which three relationships are produced by using the characteristic curves for subcritical flows at the junction \citep{ABBOTT1966,chaudhryopen}.
\subsection{Gurram model}
The Gurram model is  based on the momentum conservation principle used by \citet{Karki1997} to predict the depth ratio ($h_{1}/h_{3}$) at the junction. The equality of water depths and channel widths upstream from the junction is assumed. Here, the Gurram model is generalized to consider the discordant bottom effect and a general channel network configuration. More details about the modified Gurram formula are given in Appendix \ref{appand1}. Therefore, the unknowns at the junction can be obtained by solving the following system:
\begin{subequations}
\label{eq18}%
\begin{equation}
\sum_{k=1}^{3} \eta_{k} b_{k} h _{k} u _{k} =0
\label{eq18:1}%
\end{equation}
\begin{equation}
h_{1}=h_{2}+z_{2}
\label{eq18:2}%
\end{equation}
\begin{align}
\begin{split}
\left(\frac{h_{1}}{h_{3}}\right)^{3}\cos(\Omega)-\left(\frac{b_{3}h_{1}}{b_{1}h_{3}}\right)\Biggl[1+2{\sf{F}}^{2}-\left(\frac{2b_{2}}{b_{3}}\right)\left(\frac{h_{s}}{h_{3}^{2}}\right)z_{2}\cos(\delta)\Biggr]\\
+2{\sf{F}}^{2}\Biggl[\left(\frac{h_{1}u_{1}}{h_{3}u_{3}}\right)^{2}\cos(\Omega)+ \left(\frac{b_{3}^{2}h_{1}}{b_{1}b_{2}(h_{1}-z_{2})}\right)\left(1-\frac{b_{1}h_{1}u_{1}}{b_{3}h_{3}u_{3}}\right)\cos(\frac{8\delta} {9})\Biggr]=0
\label{eq18:3}
\end{split}%
\end{align}
\begin{equation}
u_{k} h_{k}=A_{k}h_{k}+C_{k},\quad k=1,2,3,
\label{eq18:4}%
\end{equation}
\end{subequations}
where $\sf{F}$ is the Froude number in the main downstream channel, $\Omega$ is the angle between the main upstream channel and the main downstream channel, and $\delta$ is the junction angle (the angle between the main and lateral channels); see Fig. \ref{f1}. The depth over the lateral bottom step, found using the analytical procedure by \citet{valiani2008depth} is denoted $h_{s}$. Equation (\ref{eq18:1}) represents  mass conservation, Eq. (\ref{eq18:2}) is obtained by assuming equal water elevation upstream from the junction, Eq. (\ref{eq18:3}) is the modified Gurram formula, and  Eq. (\ref{eq18:4}) represents the characteristic equations according to \citet{ABBOTT1966} and \citet{chaudhryopen}.
\subsection{Hsu model} 
The Hsu model is derived by \citet{JungLee1998}, similarly to the Gurram model, but energy and momentum  coefficients are taken into account. The unknowns at the junction are obtained by solving the following system:  
\begin{subequations}
\label{eq19}
\begin{equation}
\sum_{k=1}^{3} \eta_{k} b_{k} h _{k} u_{k}=0
\label{eq19:1}%
\end{equation}
\begin{equation}
h_{1}=h_{2}+z_{2}
\label{eq19:2}%
\end{equation}
\begin{align}
\begin{split}
\left(\frac{h_{1}}{h_{3}}\right)^{3}\cos(\Omega)-\left(\frac{b_{3}h_{1}}{b_{1}h_{3}}\right)\Biggl[1+\frac{2\beta {\sf{F}}^{2}}{\gamma}-\left(\frac{2b_{2}}{b_{3}}\right) \left(\frac{h_{s}}{h_{3}^{2}}\right)z_{2}\cos(\delta)\Biggr]\\
+\frac{2\beta {\sf{F}}^{2}}{\gamma}\Biggl[\left(\frac{h_{1}u_{1}}{h_{3}u_{3}}\right)^{2}\cos(\Omega)+ \left(\frac{b_{3}^{2}h_{1}}{b_{1}b_{2}(h_{1}-z_{2})}\right)\left(1-\frac{b_{1}h_{1}u_{1}}{b_{3}h_{3}u_{3}}\right) \cos (\delta)\Biggr]=0
\label{eq19:3}
\end{split} %
\end{align}
\begin{equation}
u_{k} h_{k}=A_{k}h_{k}+C_{k}, \quad k=1,2,3,
\label{eq19:4}%
\end{equation}
\end{subequations}
where $\beta$ is the momentum  coefficient and $\gamma$ is the energy  coefficient. Equation (\ref{eq19:1}) is the mass conservation, Eq. (\ref{eq19:2}) is obtained by assuming equal water elevation upstream from the junction, Eq. (\ref{eq19:3}) is the modified Hsu formula given in Appendix \ref{appand2}, and Eq. (\ref{eq19:4}) represents the characteristic equations according to \citet{ABBOTT1966} and \citet{chaudhryopen}.

The nonlinear systems (\ref{eq14}), (\ref{eq17}), (\ref{eq18}), or (\ref{eq19}) are solved using a hybrid iterative method \citep{powell1970hybrid}.

\section{Results for steady flows }
\label{steady}
To validate the junction models, five steady flow experiments \citep{JungLee1998,hsu1998flow,pinto2015experimental,biron1996effects,wang2007three} are numerically reproduced. Different network configurations (asymmetric and symmetric confluences) are considered, and the lateral bottom step is present at the junction in specific cases.
\subsection{Steady flow in asymmetric confluence with concordant bottom}
\begin{figure}[htpb!]
\begin{center} 
\resizebox*{14cm}{!}{\includegraphics{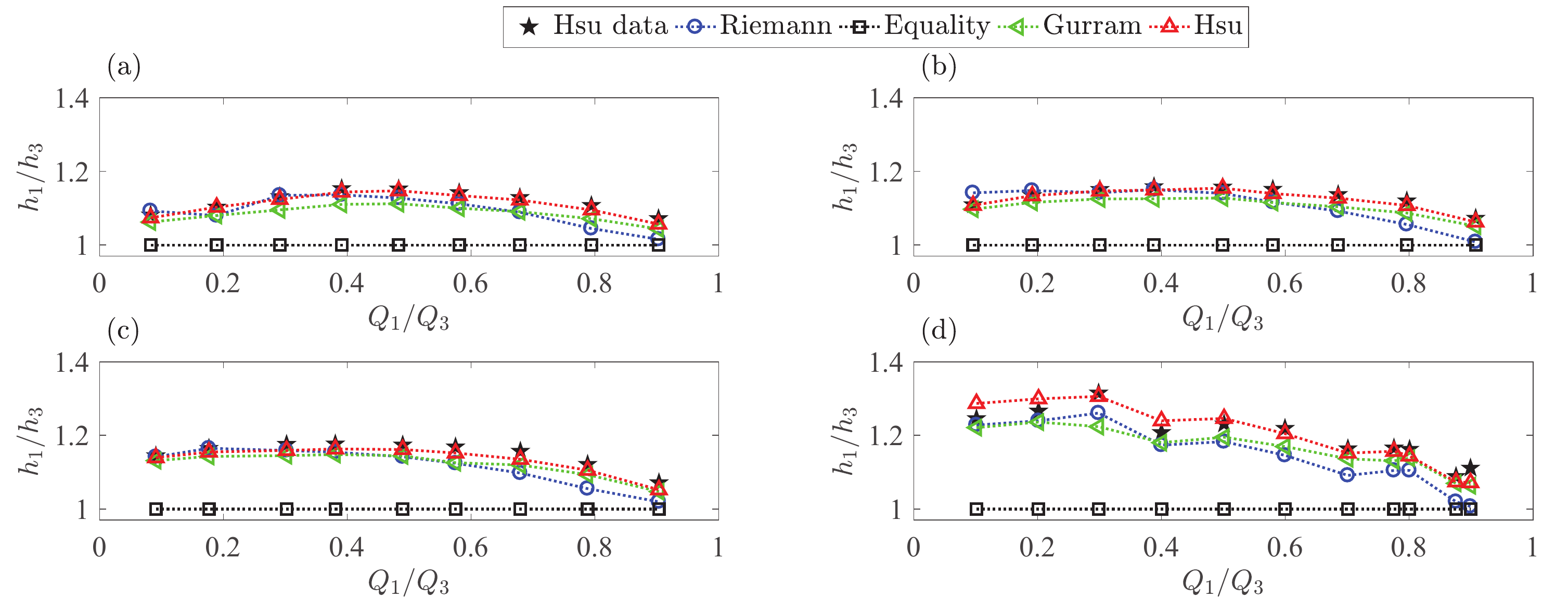}}
\caption{Different numerical solutions vs. the experimental data of \protect\citet{JungLee1998} with junction angles (a) \unit{30}{\degree} (b) \unit{45}{\degree} (c) \unit{60}{\degree}; and vs. the experimental data of \protect\cite{hsu1998flow} with  \unit{90}{\degree} junction angle (d). The experimental data are shown as filled stars; circles, squares, triangles, and diamonds indicate the RP approach, Equality model, Gurram model, and Hsu model, respectively. Note that the symbols represent the same quantities in the all following figures.}
\label{f3} 
\end{center}
\end{figure}

\begin{figure}[htpb!]
\begin{center}
\resizebox*{13cm}{!}{\includegraphics{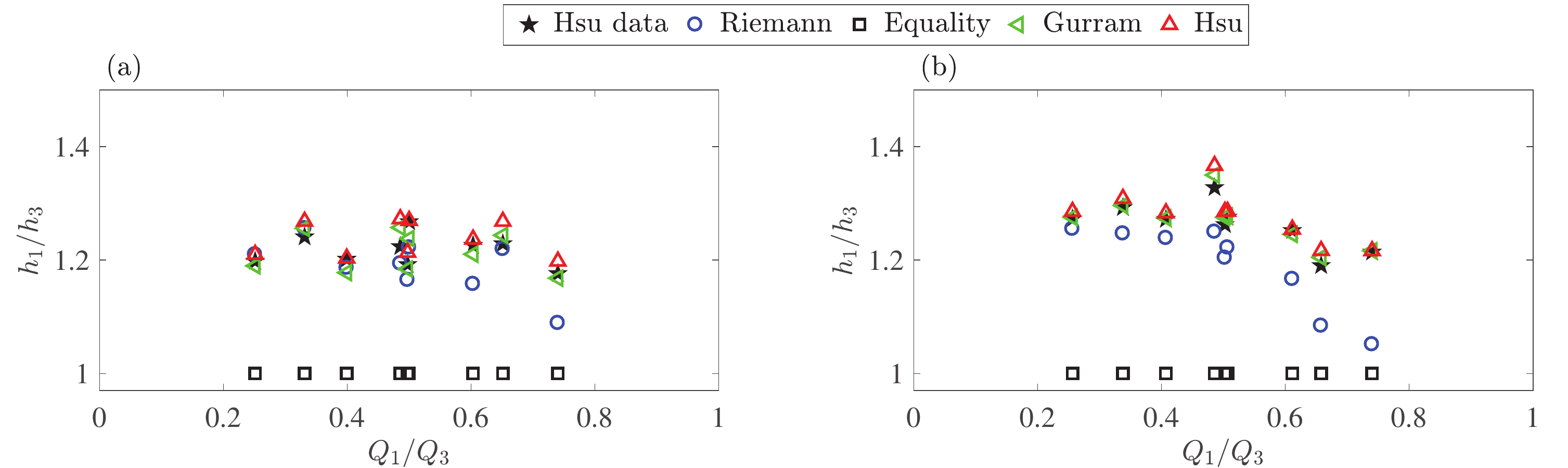}}
\caption{Different numerical solutions vs. the experimental data of \protect\citet{pinto2015experimental} with junction angles (a) \unit{30}{\degree} and (b) \unit{60}{\degree}.}
\label{f4} 
\end{center}
\end{figure}

\citet{JungLee1998} conducted experiments in a rectangular flume (Fig. \ref{f1}) with $\Omega=0$ and $z_{1}=z_{2}=z_{3}=0$. The lateral and the main channels were 1.5 and \unit{6}{\metre} long, respectively. The channel width was \unit{0.155}{\metre} for both the lateral and the main channels,  with  junction angles $\delta$ of \unit{30}{\degree}, \unit{45}{\degree}, and \unit{60}{\degree}. In \citet{hsu1998flow}, the lateral and the main channels were 4 and \unit{12}{\metre} long, respectively. The channel width was \unit{0.155}{\metre} in both channels, with a junction angle $\delta$ of \unit{90}{\degree}. In \citet{pinto2015experimental}, both channels were \unit{0.30}{\metre} wide and \unit{0.50}{\metre} deep. The main channel was \unit{10}{\metre} long, with a bottom slope of \unit{0.14}{\%} and junction angles of \unit{30}{\degree} and \unit{60}{\degree}. In this and next subsections, the values of $\beta$ and $\gamma$ are taken as 1.12 and 1.27, respectively. These values have been selected  according to the suggestions by \citet{JungLee1998}. For a quantitative comparison, the percentages of the relative error (${E}$) between the predicted depth ratio ($Y=h_{1}/h_{3}$) and the corresponding experimental values are calculated using the following formula:
\begin{equation}
\label{eq21}
{{E}} = \frac{| Y_{exp} - Y_{num}|}{Y_{exp}} \times 100,
\end{equation}
where $Y_{exp}$ refers to the experimental depth ratio (main upstream to downstream) in \citet{JungLee1998,hsu1998flow} and \citet{pinto2015experimental}, $Y_{num}$ refers to the depth ratio computed using the RP approach, Equality model, Gurram model, or Hsu model.\\
In Fig. \ref{f3}, the performance of the four junction models are compared with the data of \citet{JungLee1998,hsu1998flow}. The depth ratio $h_{1}/h_{3}$ is plotted against the discharge ratio $Q_{1}/Q_{3}$ ($Q=bhu$) with junction angles $\delta$ of \unit{30}{\degree}, \unit{45}{\degree}, \unit{60}{\degree}, and \unit{90}{\degree}. Figure \ref{f4} shows the performance of the different junction models against the \citet{pinto2015experimental} experimental data, with junction angles of \unit{30}{\degree} and \unit{60}{\degree}.
Good agreement with respect to the experiments using the RP approach, Gurram model, and Hsu model is shown. By contrast, the Equality model gives the worst behaviour, which is not surprising because such model has bad performance for {\sf{F}} greater than 0.35 \citep{kesserwani2008simulation}, and  {\sf{F}} ranges between 0.52 and 0.7 in these experiments. The percentage errors are listed in Tables \ref{tb2} and \ref{tb3}, related  to Fig. \ref{f3} and Fig. \ref{f4}, respectively.  The effect of the junction angle on the solution is clear from these Tables. Among the results, the Equality model gives the maximum  error (19.91\%) while the  Hsu model gives the minimum error (0.61\%), followed by the Gurram model (2.37\%) and RP approach (2.68\%). In general, the error of the RP approach is close to the errors of the Gurram and Hsu models for junction angles \unit{30}{\degree}, \unit{45}{\degree}, and \unit{60}{\degree}, but the difference increases for the \unit{90}{\degree} junction angle. The junction angle has a notable impact on the performance of the RP approach compared to that of the Gurram and Hsu models. Given that the RP approach does not take into account the junction angle, a reasonable motivation of this behavior can be found in the nature of the governing equations, that is, the pure shallow water equations. Neither the momentum coefficients nor energy coefficients are used, so the larger the junction angle is, the worse the agreement between the model and the real phenomenon. Clearly, the recirculation pattern becomes more important as the junction angle increases, so the performance of the RP approach can be expected to worsen as the junction angle increases.  The Gurram and Hsu models, which use empirical adjustments that take into account (more or less directly) the recirculation pattern, are less sensitive to changes in the junction angle.
A possible solution to recover the junction angle influence without empirical parameters is proposed by
\citet{bellamoli2017numerical}. In the proposed approach the junction is represented as a single two-dimensional cell connecting one-dimensional branches. 

According to this investigation, not only can the momentum-based junction models be used with acceptable error (less than 8\%, according to \citeauthor{kesserwani2008simulation}, \citeyear{kesserwani2008simulation}) but the RP approach gives tolerable errors for practical purposes. However, the use of momentum-based junction models (Gurram and Hsu models) is not trivial in many situations due to the involved empirical coefficients, such as energy and momentum coefficients, which require proper calibration based on the geometry of the junction and the characteristics of the flow dynamics.

\subsection{Steady flow in asymmetric confluence  with lateral discordant bottom} 
According to \citet{biron1996effects}, the bottom discordance has a noticeable effect on the flow in a river channel confluence, even with a small Froude number (less than 0.35). Therefore, further investigation to illustrate the behaviour of the junction models in presence of a lateral discordant bottom is presented in this subsection.  \citet{biron1996effects} performed experiments in an asymmetric channel confluence with $\Omega=0$ and $\delta=30^{\circ}$ (Fig. \ref{f1}) to investigate the effects of bottom discordance on such confluence. They describe the four flow dynamics regions at the junction, namely, the flow deflection, separation, maximum velocity, and mixing layer zones. Following the work of \citet{biron1996effects}, we consider a numerical experiment characterized by a main upstream, a lateral, and a main downstream channel, 0.12, 0.08, and \unit{0.137}{\metre} wide and 3.5, 3.5, and \unit{10}{\metre} long, respectively. The lateral bottom height is \unit{0.03}{\metre}. {\sf{F}} is less than 0.20. The discharges are $2.688\rm\times 10^{-3}$, $2.808\rm\times 10^{-3}$, and $5.496\rm\times 10^{-3}$ m$^3$s$^{-1}$ in the main upstream channel, the lateral channel, and the main downstream channel, respectively. The corresponding depths are 0.16, 0.13, and \unit{0.16}{\metre}. The discharge ratio $Q_{r}$ between the main upstream channel and the lateral channel is 1.04. The experimental data from \citet{biron1996effects} are not available. To produce cross-section averaged quantities to use as a reference solution for 1D models, TELEMAC-2D software \citep{usermanT2D} is employed. Therefore, the experiments by \citet{biron1996effects} are reproduced and the corresponding 2D numerical results are averaged on a cross section located \unit{8}{\metre} downstream from the junction.

\begin{figure}[htpb!]
\begin{center} 
\resizebox*{13cm}{!}{\includegraphics{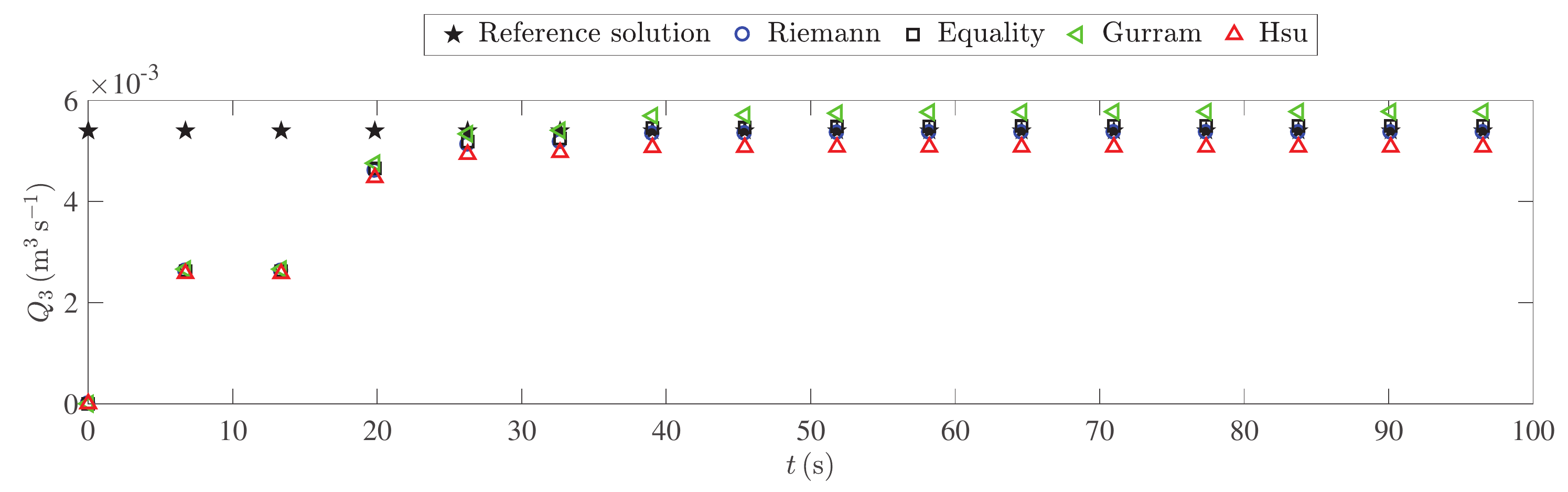}}
\caption{Four different numerical solutions for downstream discharge vs. time for a discharge ratio $Q_{r}$=1.04. Reference solution is obtained  using TELEMAC-2D software on the experimental layout of \protect\citet{biron1996effects}.}
\label{f6} 
\end{center}
\end{figure}

The behaviour of the Riemann and Equality models, which satisfactorily match the corresponding reference solution (Fig. \ref{f6}), is different from that of the Gurram model, which slightly overestimates the downstream discharge, and from that of Hsu model, that slightly underestimates the same quantity.
This difference may be due to the specific values selected for the energy and momentum  coefficients. It is worth noting that the Froude number (less than 0.35) is in the proper range of applicability of both the Gurram and Hsu models, so their complete reliability is debatable even at low Froude number. This slightly poorer performance might be due to the fact that introducing a bottom discontinuity in such methods requires a complete retuning of the empirical coefficients appearing in their formulation; these are a momentum and an energy coefficients due to the flow recirculation downstream from the junction and are tuned on the basis of flat bottom experiments: this aspect is out of the scope of the present work.

The bottom discordance divides the four models into two categories: empirical (Gurram, Hsu) and non-empirical (Riemann, Equality) models. As reported in  Table \ref{tb4}, the maximum error (7.05\%) is obtained by the Gurram model, followed by the Hsu model (5.95\%). The minimum error (0.60\%) is obtained by the RP approach, followed by the Equality model (1.78\%). 

The present computations show that even with a downstream Froude number less than 0.35, the momentum-based junction models (Gurram and Hsu) are hardly extendible to more general cases without specific studies of the role of bottom discontinuities in their physical framework. The high error of the Gurram model is very close to the 8\% limit of acceptability considered by \citet{kesserwani2008simulation}, and a certain weakness of the momentum-based methods, also for ${\sf{F}}< 0.35$, is shown. This is in contrast with the findings of \citet{kesserwani2008simulation}. Therefore, the RP approach attains the best agreement with the corresponding experimental layout of \citet{biron1996effects}.    

\subsection{Steady flow in Y-shaped confluence with lateral concordant and discordant bottoms}
\begin{figure}[htpb!]
\begin{center}
\resizebox*{13cm}{!}{\includegraphics{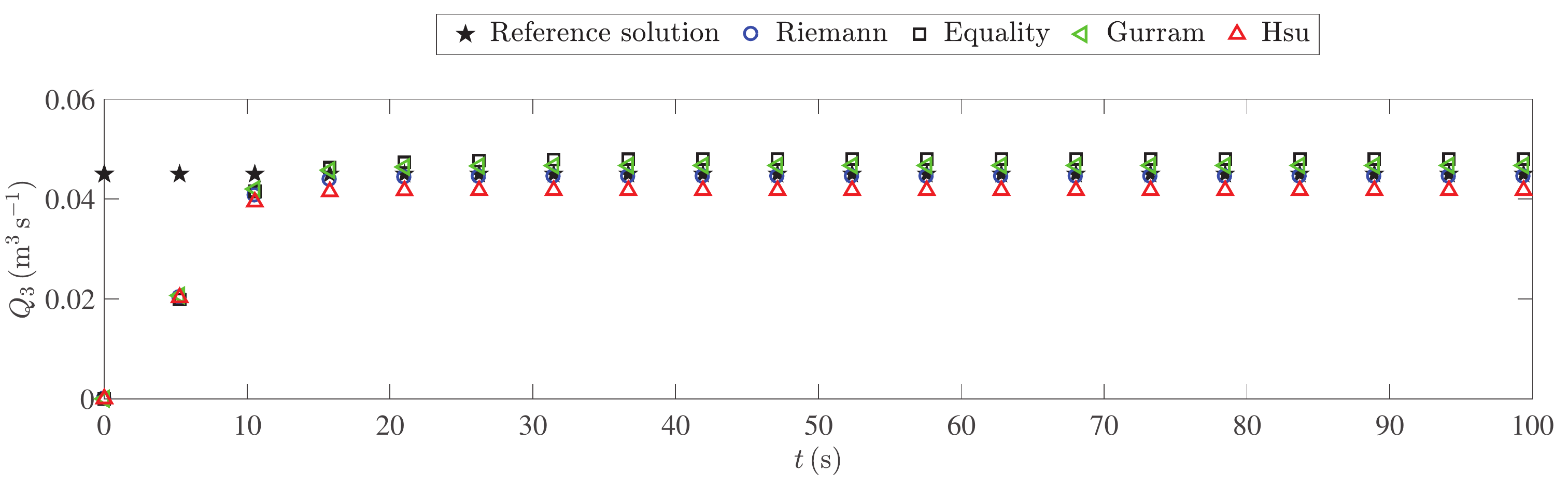}}
\caption{Different numerical solutions for downstream discharge vs. time for a discharge ratio $Q_{r}$=0.35. Reference solution is obtained  using TELEMAC-2D software on the experimental layout of \protect\citet{wang2007three}.}
\label{f8} 
\end{center}
\end{figure}

\begin{figure}[htpb!]
\begin{center}
\resizebox*{13cm}{!}{\includegraphics{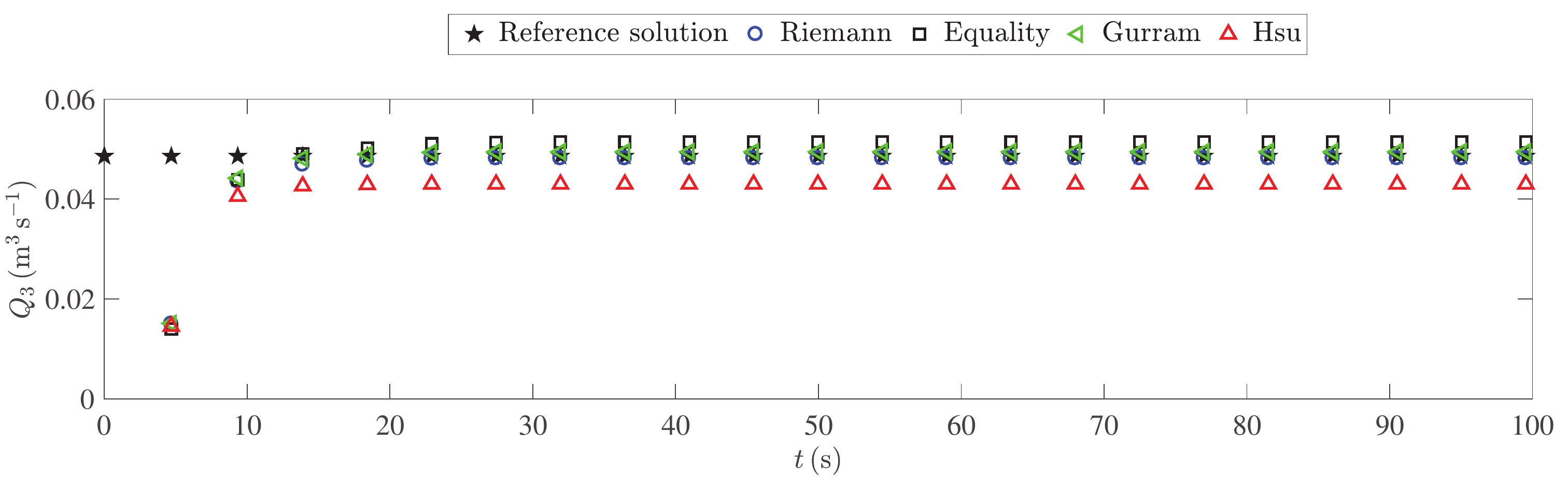}}
\caption{Four different numerical solutions for downstream discharge vs. time for a discharge ratio $Q_{r}$= 0.6. Reference solution is obtained  using TELEMAC-2D software on the experimental layout of \citet{wang2007three}.}
\label{f9} 
\end{center}
\end{figure}

The experiments performed by \citet{wang2007three} to test the effect of the bottom discordance on the flow at the Y-shaped confluence with $\Omega=\delta=45^{\circ}$ in Fig. \ref{f1} are used to compare the junction models. The \citet{wang2007three} experimental data are organized in 3D form. To use the data in a 1D framework, TELEMAC-2D software is used to reproduce the \citet{wang2007three} experimental data, and the average cross-section values of the discharge at \unit{4}{\metre} downstream from the junction are computed. Here, the lateral channel is \unit{0.3}{\metre} wide and \unit{2.4}{\metre} long; the main  upstream and downstream channels are \unit{0.45}{m} wide {and} 2.4 and \unit{4.8}{\metre} long, respectively. Two cases are considered, where the bottom is either concordant or discordant. For the concordant bottom case (i.e., $z_{2}=0$), the discharges are $3.12 \times 10^{-2}$, $1.68 \times 10^{-2}$, and \unit{4.8 \times 10^{-2}}{m^{3}s^{-1}} in the main upstream channel, the lateral channel, and the main downstream channel, respectively. The corresponding water depths are \unit{0.25}{\metre} in all channels, and the discharge ratio $Q_{r}$ between the lateral channel and the main downstream channel is 0.35. For the discordant case (i.e., $z_{2}=0.05\,\rm{m}$), the discharges are $1.8 \times 10^{-2}$, $3.0 \times 10^{-2}$, and \unit{4.8 \times 10^{-2}}{m^{3}s^{-1}}. The corresponding water depths are 0.30, 0.25, and  \unit{0.30}{\metre}, with $Q_{r}=0.6$.\\
Figures \ref{f8} and \ref{f9} compare the different junction models and the \citet{wang2007three} reference solution at the Y-shaped confluence with concordant and discordant bottom, respectively. {\sf{F}} was less than $0.27$ in both cases. However, some differences between the numerical solutions and the \citet{wang2007three} reference solution are noted. In particular, the Gurram and Equality models slightly overestimate the downstream discharge, the RP approach behaves correctly, and the Hsu model slightly underestimates the downstream discharge. The influence of the bottom on the solution can be seen in Table \ref{tb5}. The error increase by approximately 1\% when comparing the concordant and discordant bottom for the Equality model. By contrast, the error  decreases by approximately 1\%  for the Gurram model. The error increases by 9\% for the Hsu model and remains approximately constant for the RP approach.

As a conclusion of these comparisons, in the Y-shaped confluence case, the RP approach appears to outperform the momentum-based models for both concordant and discordant bottom. Indeed, some reasonable doubt arises in terms of the extent to which such momentum-based models are generalizable, especially to cases that are not strictly similar to those of the original experiments. By contrast, the RP approach, which is based on general mechanical bases, performs well, mainly with respect to case-independence. 
\section{Results for unsteady flows}
\begin{figure}[htpb!]
\begin{center}    
\resizebox*{12cm}{!}{\includegraphics{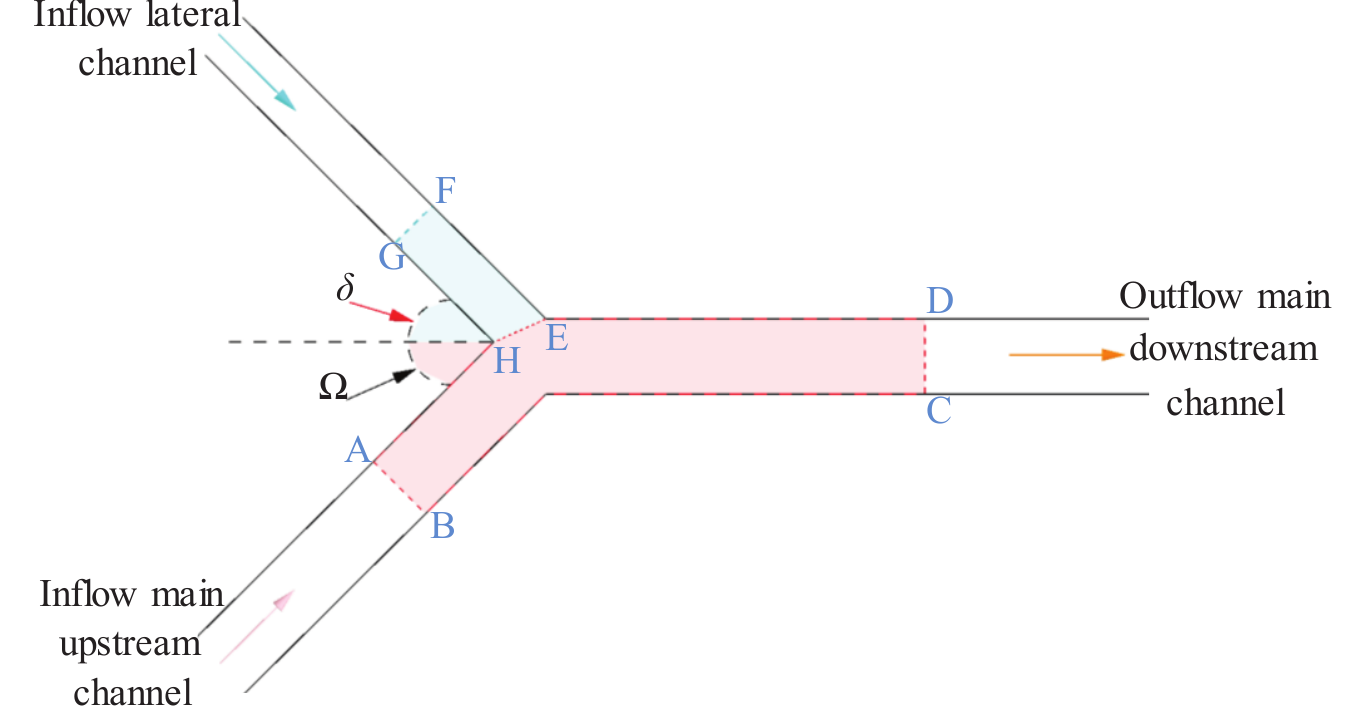}}
\caption{Star network of three channels.}
\label{f1} 
\end{center}
\end{figure}

\begin{figure}[htpb!]
\begin{center}   
\resizebox*{13cm}{!}{\includegraphics{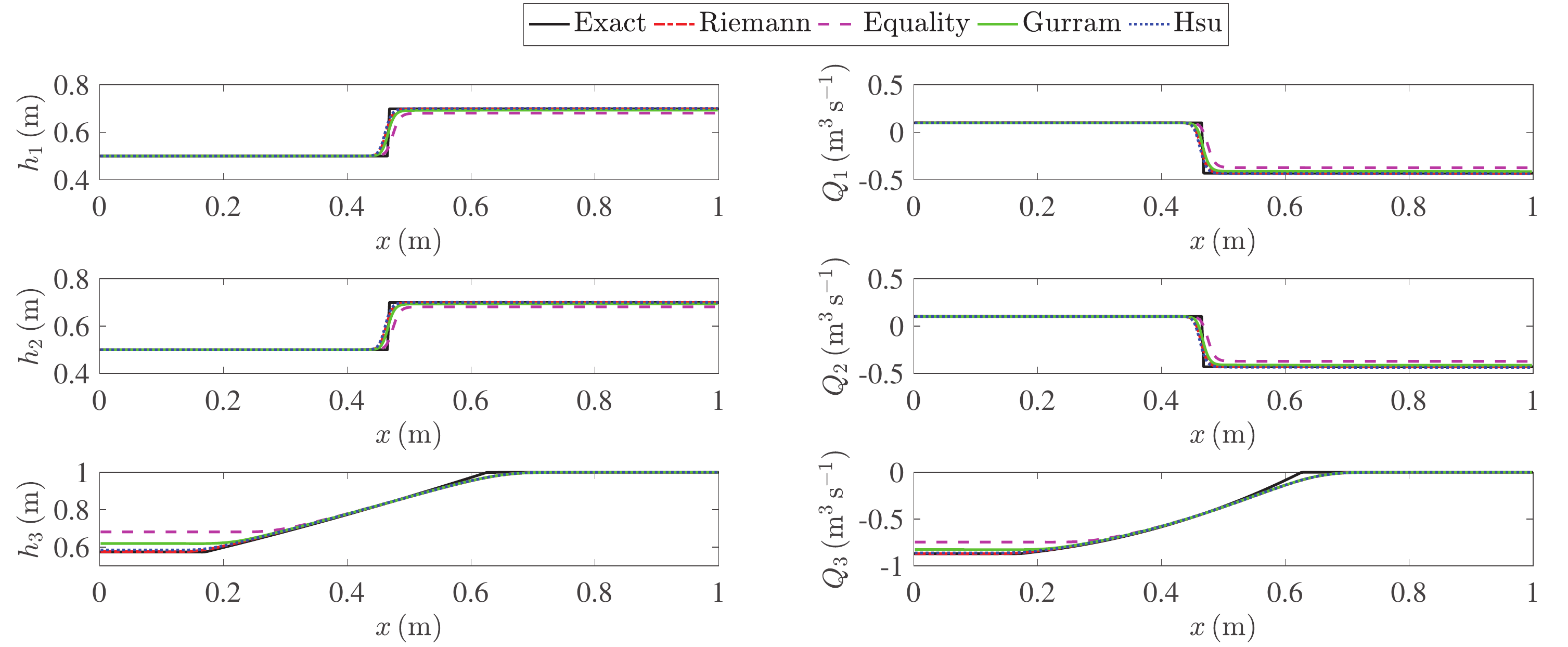}}
\caption{Water depth $h$ and discharge $Q$ evolution at time 0.2 s. The analytical solution is shown as thick solid lines; the RP approach, Equality model, Gurram model, and Hsu model are shown as dashed-dot lines, dashed lines, thin solid lines, and dots, respectively.}
\label{f10} 
\end{center}
\end{figure}

The validation of the junction models under unsteady flow conditions is not fully covered in literature. Only few studies have been performed \citep{kesserwani2008simulation,chang2015novel,briani2016notes}. Here, the analytical Riemann solution \citep{goudiaby2013riemann,JHDE2018} for unsteady flow at a junction is used to validate the four junction models. Considering the network layout shown in Fig. \ref{f1}, with $\Omega=\delta=45^{\circ}$, three channels with equal widths (i.e., $b_{1}=b_{2}=b_{3}$) and equal lengths (i.e., $L_{1}=L_{2}=L_{3}$) are connected to a single point and form a network. The discharge (and the corresponding velocity) is considered to be positive in the first and  second channels (main upstream and lateral channel) if the channel feeds the node and in the third channel (main downstream channel) if the node feeds the channel. In Fig. \ref{f1}, positive discharges correspond to arrows from left to right. The initial conditions are: $h_1=0.5\,\,\text{m}; h_2=0.5\,\,\textrm{m}; h_3=1.0\,\,\textrm{m}; Q_1=0.1\,\,\textrm{m}^{3}\textrm{s}^{-1}; Q_2=0.1\,\,\textrm{m}^{3} \textrm{s}^{-1}; Q_3=0\,\,\textrm{m}^{3}\textrm{s}^{-1}$. These conditions are chosen to obtain similar flow configurations to those of previous experimental works \citep{pinto2015experimental}. This problem is the counterpart of the dam break problem in a single channel. A shock wave travelling backward into the upstream/lateral branches and a rarefaction wave travelling forward in the downstream branch are expected. The initial state of the system (particularly, the bottom discontinuity at the junction) has important effects on the existence and uniqueness of the solution, as shown by \citet{JHDE2018}. A limited range of initial conditions allows the existence of a physically based solution. Such conditions, which are not trivial, have been derived in \citet{JHDE2018}. The current test case refers to a symmetric confluence with a continuous bottom; the bottom elevation is zero everywhere. Figure \ref{f10} shows the numerical results for the four junction models. The $l^1$ errors for the depth and the discharge are listed in Tables \ref{tb6} and \ref{tb7} and are computed according to the following formulas:
\begin{subequations}
\begin{empheq}{align}
e_{k}^{h}=\Delta x \sum_{i=1}^{N}|h^{*}_{k}(x_{i},t)-h_{k}(x_{i},t)|,\quad k=1,2,3\label{eq20:1}%
\\
e_{k}^{Q}=\Delta x \sum_{i=1}^{N}|Q^{*}_{k}(x_{i},t)-Q_{k}(x_{i},t)|,\quad k=1,2,3,\label{eq.20:2}%
\end{empheq}
\label{eq20}%
\end{subequations}
where $h^{*}$ and $Q^{*}$ are the depth and the discharge obtained using the analytical solution. $h$ and $Q$ are the depth and discharge computed by the FVM-DOT model including the junction models at the final time ($t=0.2$ s). $N$ is the number of mesh cells. It is clear that the numerical solution based on the RP approach has the best performance in this case because the only difference between that solution and the analytical one is just the numerical error. Therefore, this test is mainly devoted to understanding the performance of the classic models. The results for the shock backward propagation in the main and lateral branches are quite good for all models, with slightly worse behavior for the Equality model. In the downstream main channel, the behavior of Equality model is again the worst, followed by the Gurram model, whilst the Hsu model performs well for both depth and discharge.
\section{Conclusions}
\label{conclusion}
In this research, the use of a suitable RP approach to set up the internal boundary conditions at the junctions in the numerical simulation of channel network flows is evaluated. Generally, the RP approach matches the experimental data, despite of the geometric characteristics of the junction. Moreover, this study confirms the poor performance of the Equality model. The junction angle has a notable impact on the performance of the RP approach compared to the Gurram and Hsu models when fitting experimental data concerning the asymmetric confluence of channels with equal widths and concordant bottoms ({\sf{F}} ranges from 0.5 to 0.7).  By contrast, for the asymmetric confluence of channels with non-equal widths and with lateral discordant bottoms and for Y-shaped confluence, the Gurram and Hsu models differ substantially from the reference experimental data, even for {\sf{F}} smaller than 0.35, and the RP approach performs better. For unsteady flows, the presented results show that the RP approach has the best agreement with the analytical solutions. Therefore, the RP approach proves to be generally a good choice and has the following benefits: the approach is based on a theoretical background that is generalizable and does not rely on empirical coefficients; the overall behavior is generally satisfactory, both for steady and unsteady flows.

There are also limitations to the applicability of the RP approach. The RP approach is not validated for junctions in meandering rivers and curved channels. Moreover, recirculation and turbulence phenomena (detachment of vortexes and three-dimensional effects) are not taken into account, but this is not considered to be a severe drawback when studying problems at longitudinal scales much larger than the channel width.

\begin{table}[htpb!]
\caption{The error percentage in the computed depth ratio $h_{1}/h_{3}$  at the junction, compared to the experimental data of \protect\citet{JungLee1998,hsu1998flow}.}
{ 
\begin{tabular}[c]{@{}lcccc}\toprule
\textsl{\rm{Junction angle} $\delta$}&\rm{Riemann}&\rm{Equality }&\rm{Gurram} &\rm{Hsu}\\
\midrule
\unit{30}{\degree} &2.68&10.59&3.02&0.72\\
\unit{45}{\degree} &2.87&11.64&2.37&0.61\\
\unit{60}{\degree} &2.88&13.02&2.48&1.27\\
\unit{90}{\degree} &5.84&19.91&3.78&2.21\\
\bottomrule
\end{tabular}}
\label{tb2}
\end{table}
\begin{table}[htpb!]
\caption{The error percentage in the computed depth ratio $h_{1}/h_{3}$  at the junction, compared to the experimental data of \protect\citet{pinto2015experimental}.} 
{ 
\begin{tabular}[c]{@{}lcccc}\toprule
\textsl{\rm{Junction angle} $\delta$}&\rm{Riemann}&\rm{Equality }&\rm{Gurram} &\rm{Hsu}\\
\midrule
\unit{30}{\degree} &2.83&17.84&1.54&1.64\\
\unit{60}{\degree} &5.62&20.63&0.70&1.49\\
\bottomrule
\end{tabular}}
\label{tb3}
\end{table}
\begin{table}[htpb!]
\caption{The error percentage in the computed downstream discharge $Q_{3}$ relative to the reference solution obtained  using TELEMAC-2D software on the experimental layout of \protect\citet{biron1996effects}.} 
{
\begin{tabular}[c]{@{}lc}\toprule
\rm{Junction model}&\rm{Discordant bottom}\\
\midrule
\rm{Riemann } &0.59\\
\rm{Equality } &1.77\\
\rm{Gurram}& 7.05\\
\rm{Hsu}& 5.95 \\
\bottomrule
\end{tabular}}
\label{tb4}
\end{table}
\begin{table}[htpb!]
\caption{The error percentage in the computed downstream discharge $Q_{3}$ relative to the reference solution obtained  using TELEMAC-2D software on the experimental layout of \protect\citet{wang2007three}.} 
{
\begin{tabular}[c]{@{}lcc}\toprule
\rm{Junction model}&\rm{Concordant bottom}&\rm{Discordant bottom}\\
\midrule
\rm{Riemann } &{1.079} &{{1.25}}\\
\rm{Equality } &{6.667} & {8.27}\\
\rm{Gurram}&{3.858}  &{2.53}\\
\rm{Hsu}&{7.223} &{16.21}\\
\bottomrule
\end{tabular}}
\label{tb5}
\end{table}
\begin{table}[htpb!]
\caption{The $l^1$ error in the computed depth. Analytical solution by  \protect\citet{goudiaby2013riemann}.} 
{ 
\begin{tabular}[c]{@{}lccc}\toprule
\rm{Junction model}&\rm{Upstream main channel}&\rm{Lateral channel}&\rm{Downstream main channel}\\
\midrule
\rm{Riemann\rule{0ex}{2.6ex}}   &1.6997$\times 10^{-3}$ &1.6997$\times 10^{-3}$ & 4.2037$\times 10^{-3}$ \\
\rm{Equality } &1.1143$\times 10^{-2}$ & 1.1143$\times 10^{-2}$ & 2.8994$\times 10^{-2}$\\
\rm{Gurram}&4.3977$\times 10^{-3}$  &4.3977$\times 10^{-3}$& 1.2914$\times 10^{-2}$\\
\rm{Hsu }&2.1425$\times 10^{-3}$ &2.1425$\times 10^{-3}$& 5.9129$\times 10^{-3}$\\
\bottomrule
\end{tabular}}
\label{tb6}
\end{table}
\begin{table}[htpb!]
\caption{The $l^1$ error in the computed discharge. Analytical solution by \protect\citet{goudiaby2013riemann}.} 
{ 
\begin{tabular}[c]{@{}lccc}\toprule
\rm{Junction model}&\rm{Upstream main channel}&\rm{Lateral channel}&\rm{Downstream main channel}\\
\midrule
\rm{Riemann\rule{0ex}{2.6ex}}  &4.2644$\times 10^{-3}$ &4.2644$\times 10^{-3}$ &8.8333$\times 10^{-3}$ \\\rm{Equality } &3.5669$\times 10^{-2}$ &3.5669$\times 10^{-2}$ &3.8996$\times 10^{-2}$\\
\rm{Gurram}    &1.1279$\times 10^{-2}$ &1.1279$\times 10^{-2}$ &1.7570$\times 10^{-2}$\\
\rm{Hsu }      &6.4912$\times 10^{-3}$ &6.4912$\times 10^{-3}$ &1.0272$\times 10^{-2}$\\
\bottomrule
\end{tabular}}
\label{tb7}
\end{table}
\appendix
\section*{Funding}
The first author is grateful to the Egyptian Ministry of Higher Education (MOHE: 2014/2018) and to the National Authority for Remote Sensing and Space Sciences (NARSS) for financial support. The third author is funded by the University of Ferrara within the Founding Program FIR 2016, project title "Energy-preserving numerical models for the Shallow Water Equations".
\section{The modified Gurram formulas}
\label{appand1}
In this appendix, we introduce the modified Gurram formula. We consider the junction in Fig.~\ref{f1}.  Following the assumptions made by \citet{Karki1997}, the flow is assumed to be steady with small bottom slope, such that the friction slope is nearly compensated; the flow is one-dimensional in the main upstream and downstream channels with momentum and energy  coefficients ($\beta_{1}$ at AB, $\beta_{3}$ at CD, and $\gamma$ at CD) assumed to be unity. Towards the junction, the lateral channel flow is accelerated due to the flow contraction at the separation zone \citep{Karki1997}. Therefore, the lateral channel momentum $M_{2}$ at EH in Fig.~\ref{f1} can be written as  
\begin{equation}
\label{aeq1}
M_{2}=\beta_{2}\rho b_{2}h_{2}u_{2}^{2} \cos(\delta),
\end{equation}
where $\rho$ is the density of water and $\beta_{2}$ is the lateral momentum coefficient. \citet{Karki1997} showed that $\beta_{2}$ can be computed with the following relation
\[ \beta_{2}=\frac{b_{3}h_{3}u_{3}}{b_{2}h_{2}u_{2}}\frac{\cos(\alpha)}{\cos(\delta)}, \]
 where $\alpha$ as the angle between the representative lateral velocity vector at EH and the main channel direction. The relation between $\alpha$ and $\delta$ is  \citep{hager1987discussion} \[\alpha=\frac{8}{9}\delta.\]
Because of the presence of the bottom step in the lateral channel, the flow mixing between the upstream and the lateral channel is expected to be increased; therefore, the relationship between $\delta$ and $\alpha$ must be recalibrated. However, this process is beyond the scope of this work, so the suggestion of \citet{hager1987discussion} is maintained. Assuming a hydrostatic pressure distribution, the force exerted  by the lateral bottom step is computed according to \citet{JHDE2018}, and the equality of the water level between the upstream main channel and the lateral channel is assumed rather than the equality of water depth. Taking into account the angle $\Omega$, the momentum balance in the main downstream channel direction over a control volume ABCDEH in Fig.~\ref{f1} gives
\begin{equation}
\begin{split}
\label{aeq4}
 \rho b_{1} h_{1}u_{1}^{2}\cos(\Omega)+\frac{\rho g}{2}b_{1}h^{2}_{1}\cos(\Omega)+ \rho b_{3}h_{3}u_{3}u_{2} \cos(\alpha)+\rho g b_{2}h_sz_{2}\cos(\delta)=\\
 \rho b_{3}h_{3} u_{3}^{2} +\frac{\rho g }{2}b_{3}h^{2}_{3},
\end{split}
\end{equation}  
where $h_{s}$ refers to the depth over the lateral bottom step, which is computed  by applying the conservation of the total head over the step \citep{valiani2008depth} by considering only the subcritical solution
\begin{equation}
\label{aeq5}
h_{s}=\frac{1}{3}\left(h_{3}+\frac{u_{3}^{2}}{2g}-z_{2}\right)\left[1-2\cos\left(\frac{2\pi +\theta}{3}\right)\right]
\end{equation}
where \[\theta=\arccos\left(1-27\left( \frac{-h_{3}-\frac{u_{3}^{2}}{2g}+z_{2}}{\left(\frac{h_{1}u_{1}}{h_{3}u_{3}}\right)^{2}\left( \frac{h_{3}^{2}u_{3}^{2}}{g}\right)^{\frac{1}{3}}}\right) \right). \]
For more details, see \citep{JHDE2018}. Multiplying Eq. (\ref{aeq4})  by $2/b_{3}h_{3}^{2}$ gives
\begin{multline}
\label{aeq6}
\left[\frac{2b_{1}h_{1}u_{1}^{2}}{gb_{3}h_{3}^{2}}+\left(\frac{b_{1}}{b_{3}} \right)\left(\frac{h_{1}}{h_{3}}\right)^{2}\right]\cos(\Omega)+ \frac{2u_{2}u_{3}}{g h_{3}}\cos(\alpha)+\left(\frac{2b_{2}}{b_{3}}\right)\left(\frac{h_{s}}{h_{3}^{2}}\right)z_{2}\cos(\delta)= \frac{2u_{3}^{2}}{gh_{3}}+1.
\end{multline}
The continuity equation implies
\begin{equation}
\label{aeq66}
b_{1}h_{1}u_{1}+b_{2}h_{2}u_{2}=b_{3}h_{3}u_{3}.
\end{equation} 
By using the equality of the water level upstream from the junction and substituting Eq. (\ref{aeq66}) into Eq. (\ref{aeq6}), with a little arrangement we obtain
\begin{multline}
\label{aeq7}
\left(\frac{h_{1}}{h_{3}}\right)^{3}\cos(\Omega)-\left(\frac{b_{3}h_{1}}{b_{1}h_{3}}\right)\Biggl[1+2{\sf{F}}^{2}-\left(\frac{2b_{2}}{b_{3}}\right)\left(\frac{h_{s}}{h_{3}^{2}}\right)z_{2}\cos(\delta)\Biggr]+ 2{\sf{F}}^{2}\Biggl[\left(\frac{h_{1}u_{1}}{h_{3}u_{3}}\right)^{2}\cos(\Omega)+\\ \left(\frac{b_{3}^{2}h_{1}}{b_{1}b_{2}(h_{1}-z_{2})}\right)\left(1-\frac{b_{1}h_{1}u_{1}}{b_{3}h_{3}u_{3}}\right) \cos(\frac{8\delta}{9})\Biggr]=0,
\end{multline}
where \[{\sf{F}}=\sqrt{\frac{\gamma u^{2}_{3}}{gh_{3}}};\] therefore, Eq. (\ref{aeq7}) represents the final Gurram formula that is used in the model (\ref{eq18}).
\section{The modified Hsu formulas}
\label{appand2}
This appendix shows the derivation of the modified Hsu formula in the channel network of Fig.~\ref{f1}. According to \citet{hsu1998flow}, the flow is accelerated due to the flow contraction at the separation zone as long as we move towards the junction. Therefore, momentum coefficients are introduced ($\beta_{1}$ at AB, $\beta_{2}$ at FG, $\beta_{3}$ at CD, and $\beta_{{EH}}$ at EH). Assuming steady flow and a hydrostatic pressure distribution, neglecting the friction force, taking into account the  angle $\Omega$, the acting force due to the presence of the lateral step, and further assuming $\beta_{1}=\beta_{2}=\beta_{3}=\beta_{{EH}}=\beta$, the momentum balance in the main downstream channel direction over the area ABCDEH gives
\begin{equation}
\begin{split}
\label{beq2}
\beta \rho b_{1}h_{1}u_{1}^{2}\cos(\Omega)+\frac{\rho g}{2}b_{1}h^{2}_{1}\cos(\Omega)+\beta\rho b_{2}h_{2}u_{2}u_{{EH}} \cos(\alpha)+\rho g b_{2}h_sz_{2}\cos(\delta)=\\
\beta \rho b_{3}h_{3}u_{3}^{2} +\frac{\rho g }{2}b_{3}h^{2}_{3},
\end{split}
\end{equation}  
where $u_{{EH}}$ is the representative velocity at EH. According to \citet{hsu1998flow}, the representative velocity $u_{{EH}}$ is related to the angle $\alpha$  by
\begin{equation}
\label{beq3}
u_{{EH}}=\frac{b_{2}h_{2}u_{2}}{b_{{EH}}h_{{EH}}\sin(\alpha)}.
\end{equation}  
$b_{{EH}}$ and $h_{{EH}}$ are the channel width and the water depth at section EH in the lateral channel, respectively. Substituting Eq. (\ref{beq3})  into Eq. (\ref{beq2})  gives
\begin{equation}
\begin{split}
\label{beq5}
\beta\rho b_{1}h_{1}u_{1}^{2}\cos(\Omega)+\frac{\rho g }{2}b_{1}h^{2}_{1}\cos(\Omega)+\frac{\beta \rho (b_{2}h_{2}u_{2})^{2}}{b_{{EH}}h_{{EH}}} \cot(\alpha)+ b_{2}h_sz_{2}\cos(\delta)=\\
\beta\rho b_{3}h_{3}u_{3}^{2} +\frac{ \rho g}{2}b_{3}h^{2}_{3}.
\end{split}
\end{equation} 
Applying  the momentum balance in the lateral channel direction over the area EFGH gives
\begin{equation}
\label{beq6}
 \beta\rho b_{2}h_{2} u_{2}^{2}+\frac{\rho g}{2}b_{2}h^{2}_{2}=\frac{\rho g}{2}b_{2}h^2_{{EH}}+  \frac{\beta \rho (b_{2}h_{2}u_{2})^{2}}{b_{{EH}}h_{{EH}}}\frac{\cos(\delta-\alpha)}{\sin(\alpha)}.
\end{equation}
Taking into account the equality of {the} water level upstream from the junction, letting $b_{{EH}}=b_{2}/\sin(\delta)$, further assuming $h_{{EH}}=h_{2}$ based on experimental observation by \citet{hsu1998flow}, taking into account the effect of the lateral bottom step $h_{s}$ \citep{valiani2017momentum}, using the mass continuity equation, and substituting Eq. (\ref{beq6}) into Eq. (\ref{beq5}), it possible to obtain
\begin{multline}
\label{beq8}
\left(\frac{h_{1}}{h_{3}}\right)^{3}\cos(\Omega)-\left(\frac{b_{3}h_{1}}{b_{1}h_{3}}\right)\Biggl[1+\frac{2\beta {\sf{F}}^{2}}{\gamma}-\left(\frac{2b_{2}}{b_{3}}\right)\left(\frac{h_{s}}{h_{3}^{2}}\right) z_{2}\cos(\delta)\Biggr]+ \frac{2\beta {\sf{F}}^{2}}{\gamma}\Biggl[\left(\frac{h_{1}u_{1}}{h_{3}u_{3}}\right)^{2}\cos(\Omega)+ \\ \left(\frac{b_{3}^{2}h_{1}}{b_{1}b_{2}(h_{1}-z_{2})}\right)\left(1-\frac{b_{1}h_{1}u_{1}}{b_{3}h_{3}u_{3}}\right) \cos (\delta)\Biggr]=0.
\end{multline}
Therefore, Eq. (\ref{beq8}) represents the modified Hsu formula (\ref{eq19:3}).
\section*{Notation}
\begin{tabular}{@{}ll}
$\mathbf{A}$&= Jacobian matrix of the flux function  (--)\\
$b_{k}$&= channel width of the $k$th channel (m)\\
$\rm{CFL}$&= the Courant-Fredrich-Lewy coefficient (--)\\ 
$c$&= wave celerity (ms$^{-1}$)\\
$\bm{D}$&= fluctuation term (--)\\
${E}$&= relative percent error (--)\\
$e_{k}^{h}$&= error in the depth of the $k$th channel (--)\\
$e_{k}^{Q}$&= error in the discharge of the $k$th channel (--)\\
$\sf{F}$& = downstream Froude number (--)\\
$\bm{F}$&= flux function (--)\\
${g}$&= gravity acceleration (ms$^{-2}$)\\
$h_{k}$&= water depth in the $k$th channel (m)\\
$h^{*}_{k}$&= depth (analytical solution) in the $k$th channel (m)\\
$h_{s}$&= depth over the lateral bottom step (m)\\
$k$&= channel index, $k=1$ refers to the main upstream channel,\\
    &\quad $k=2$ refers to the lateral channel, $k=3$ refers to the main\\
    &\quad downstream channel (--)\\
$L_{k}$&= channel length of the  $k$th channel (m)\\
$M_{k}$&= momentum in the $k$th channel (\rm{N})\\
$N$&= number of mesh cells (--)\\
$n$&= time step index (--)\\
$Q_{k}$&= water discharge in the $k$th channel (m$^3$s$^{-1}$)\\
$Q_{r}$&= discharge ratio (--)\\
$Q^{*}_{k}$&= discharge (analytical solution) in the $k$th channel (m$^3$s$^{-1}$)\\
$\bm{S}$&= vector of source term (--)\\
$S_{0x}$&= bottom slope (--)\\
$S_{f}$&= friction slope (--)\\
$s$&= parameter, $s\in [0,1] $ (--) \\
$t$&= time (s)\\
$\bm{U}$&= vector of conservative variables (--)\\
$u_{k}$&= water velocity in the $k$th channel (ms$^{-1}$)\\
$\bm{W}$&= extended vector of conservative variables (--)\\
$\bm{W}_{i}^n$&= space average of $\bm{W}$ over the $i$th cell at time $t^n$ (--)\\
$x_{k}$&= space in the $k$th channel (m)\\
$Y_{exp}$&= experimental depth ratio (main upstream to downstream) (--)\\
$Y_{num}$&= computed depth ratio (main upstream to downstream) (--)\\
$z_{k}$&= bottom elevation in the $k$th channel (m)\\
$\alpha$&= angle between the lateral velocity vector at EH \\
         &\quad and the main channel direction (--)\\
$\beta$&= momentum  coefficient, $\beta=1.27$ (--)\\
$\gamma$&= energy  coefficient, $\gamma=1.12$ (--)\\
$\delta$&= junction angle (--)\\
$\eta_{k}$&= parameter number in the $k$th channel, $\eta=-1,1$  (--)\\
$\rho$&= water density (kgm$^{-3}$)\\ 
$\mathbf{\psi}$&= integral path (--)\\
$\Omega$ &= main channel angle (--)
\end{tabular} 
\bibliographystyle{apacite}


\end{document}